\documentclass[iop]{emulateapj}
\usepackage{apjfonts}
\newcommand{\be}{\begin{equation}}
\newcommand{\ee}{\end{equation}}

\newcommand\ionn[2]{#1$\;${\scshape{#2}}}
\def\kms{\ {\rm km\, s}^{-1}}

\shortauthors{CONROY, VAN DOKKUM, \& GRAVES}
\shorttitle{Strontium and Barium In Early-Type Galaxies}

\begin{document}

%----------------------------------------------------------------
\title{Strontium and Barium In Early-Type Galaxies}
%----------------------------------------------------------------

\author{Charlie Conroy\altaffilmark{1},
  Pieter G. van Dokkum \altaffilmark{2}, 
  and Genevieve J. Graves\altaffilmark{3,4}}

\altaffiltext{1}{Department of Astronomy \& Astrophysics, University
  of California, Santa Cruz, CA, USA}
\altaffiltext{2}{Department of Astrophysical
  Sciences, Yale University, New Haven, CT, USA}
\altaffiltext{3}{Department of Astronomy, University of California, 
  Berkeley, CA, USA}
\altaffiltext{4}{Department of Astrophysical Sciences, Princeton
  University, Princeton, NJ, USA}

\slugcomment{accepted for publication in ApJ Letters}

\begin{abstract}

  The detailed abundance patterns of the stars within galaxies provide
  a unique window into the history of star formation (SF) at early
  times.  Two widely used `chronometers' include the $\alpha$ and
  iron-peak elements, which are created on short and long timescales,
  respectively.  These two clocks have been widely used to estimate SF
  timescales from moderate-resolution spectra of early-type galaxies.
  Elements formed via $s$-process neutron captures (e.g., Sr and Ba)
  comprise a third type of chronometer, as the site of the main
  $s$-process is believed to be intermediate and low-mass asymptotic
  giant branch stars.  The [$\alpha$/Ba] ratio in particular should
  provide a powerful new constraint on the SF histories of galaxies,
  in part because it is insensitive to the uncertain distribution of
  Type Ia SNe detonation times and the overall Ia rate. Here we
  present new measurements of the abundance of Sr and Ba in nearby
  early-type galaxies by applying stellar population synthesis tools
  to high S/N optical spectra. We find a strong anti-correlation
  between [Mg/Fe] and [Ba/Fe], and a strong positive correlation
  between [Mg/Ba] and galaxy velocity dispersion.  These trends are
  consistent with the idea that more massive galaxies formed their
  stars on shorter timescales compared to less massive galaxies, and
  rule out several other proposed explanations for the observed
  super-solar [Mg/Fe] values in massive galaxies.  In contrast,
  [Sr/Fe]\,$\sim0$, with no strong variation across the sample.  It is
  difficult to interpret the Sr trends without detailed chemical
  evolution models owing to the multiplicity of proposed
  nucleosynthetic sites for Sr.

\end{abstract}


%-----------------------------------------------------------------%

\section{Introduction}
\label{s:intro}

While it is clear that the majority of massive early-type galaxies
formed the bulk of their stars at $z>1$, the precise nature and
duration of star formation in these galaxies has proven difficult to
constrain.  The difficulty arises from the fact that the main sequence
turnoff point varies little with age at late times, implying that the
integrated spectral energy distribution varies little with age at late
times.  For old stellar systems such as early-type galaxies, it has
become common to instead rely on elemental abundance ratios to probe
the star formation histories at early times \citep[e.g.,][]{Worthey92,
  Thomas05}.  The classic example is the ratio of $\alpha$ to
iron-peak elements,
[$\alpha$/Fe]\footnote{[X/Fe]$\equiv$log(X/Fe)$-$log(X/Fe)$_\odot$}.
The $\alpha$ elements form mostly in massive stars, which evolve on
short timescales, while iron-peak elements form mostly in Type Ia
supernovae (SNe), which occur on longer timescales.  The [$\alpha$/Fe]
ratio is thus a sensitive probe of SF on $\sim10^{8-9}$ yr timescales
\citep[e.g.,][]{Tinsley79, Thomas98}.

One of the principal difficulties in using [$\alpha$/Fe] to measure a
SF timescale is the unknown Type Ia SNe delay time distribution. Other
difficulties include the possibility that the overall Ia rate varies
in some systematic way with galaxy properties, the possibility of
selective mass-loss such that Fe is preferentially lost from the
system, and potential variation in the initial mass function
\citep{Worthey92, Thomas99, Trager00}.  A promising alternative
chronometer is Ba, which is believed to form predominately within the
envelopes of asymptotic giant branch stars via $s$-process neutron
captures \citep{Burbidge57, Busso99, Herwig05}.  The precise
mass-dependent yields of Ba are presently not well-known, but there is
hope that progress on this topic can be made in the near-term (in
contrast to the delay time distribution for Ia's, which appears to be
governed largely by the initial conditions of stellar binarity).  The
prospects for measuring Ba in the integrated light spectra of galaxies
are favorable owing to the strong transitions of \ionn{Ba}{ii} in the
blue spectral region.

Sr is another neutron capture element with strong transitions in the
blue.  Like Ba, Sr is predominantly produced by the $s$-process, at
least in the solar system.  These two elements probe two of the three
$s$-process peaks, with Sr belonging to the first (along with Y and
Zr), and Ba belonging to the second (along with La, Ce, Pr, and Nd).
While the nucleosynthetic origin of Ba is relatively secure, the same
cannot be said for the elements in the first $s$-process peak
\citep{Couch74, Woosley92, Raiteri93, Sneden08}.  Indeed, current
chemical evolution models of the Galaxy are unable to reproduce the
observed behavior of Sr-Y-Zr at low metallicity without appealing to
exotic and/or ad hoc nucleosynthetic sites
\citep[e.g.,][]{Travaglio04, Qian08}.  Perhaps the [Sr/Ba] ratios in
metal-rich massive galaxies will provide some insight into this
problem.

In this Letter we employ our new stellar population synthesis (SPS)
model to interpret high quality optical spectra of early-type galaxies
in order to constrain the abundances of Sr and Ba.  In Sections
\ref{s:model} and \ref{s:data} we describe the model and data, and in
Section \ref{s:res} we present our results.

%-----------------------------------------------------------------%

\begin{figure}[!t]
\center
\resizebox{3.5in}{!}{\includegraphics{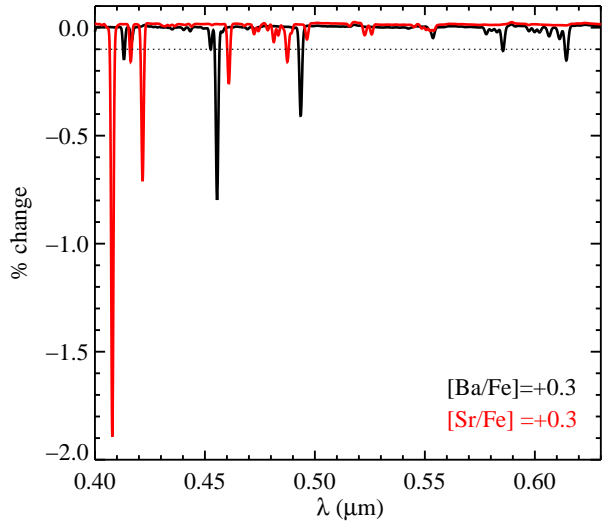}}
\caption{Change in the spectrum due to a 0.3 dex enhancement in the
  [Sr/Fe] and [Ba/Fe] abundance ratios.  The model is an integrated
  light spectrum with an age of 13 Gyr, solar metallicity, and a
  \citet{Kroupa01} IMF.  The spectra have been smoothed with a
  velocity dispersion of $\sigma=350\kms$, which is the same smoothing
  scale applied to the SDSS data.  The dotted line represents the
  sensitivity limit for a signal-to-noise ratio of S/N$=10^3$, which
  is the typical S/N for the SDSS stacks in this wavelength range.}
\label{fig:response}
\end{figure}

\section{Model}
\label{s:model}

The SPS model used herein was developed in \citet[][CvD12]{Conroy12a}.
The model adopts standard SPS techniques, including constructing
libraries of isochrones and stellar spectra.  Empirical spectra form
the core of the model.  The empirical stars are of approximately solar
metallicity and have solar abundance patterns.  We computed a large
grid of model stellar atmospheres and spectra in order to construct
response functions (i.e., the relative change in the spectrum of a
star due to a change in the abundance of a single element).  The model
atmospheres and spectra were computed with the ATLAS model atmosphere
and spectrum synthesis package \citep{Kurucz70, Kurucz93}, ported to
Linux by \citet{Sbordone04}.  Specifically, we use the ATLAS12 code,
computing new atmospheres for each change in abundance pattern.  The
line list was provided by
R. Kurucz\footnote{\texttt{kurucz.harvard.edu/}}, including linelists
for TiO and H$_2$O, amongst other molecules.  The spectral response
functions were applied to the empirical stellar spectra in order to
create models with arbitrary abundance patterns.  The model allows for
arbitrary variation in the initial mass function (IMF) and spans ages
from $3-13.5$ Gyr.  See CvD12 for further details regarding the model.

We follow \citet{Conroy12b} in fitting the model to data.  In its
present form the model contains 27 free parameters, including the
redshift and velocity dispersion, a two-part power-law IMF, two
population ages, four nuisance parameters, and the abundances of C, N,
Na, Mg, Si, Ca, Ti, V, Cr, Mn, Fe, Co, Ni, Sr, Y, and Ba, and O,Ne,S
are varied in lock-step.  These parameters are fit to the data via a
Markov Chain Monte Carlo fitting technique.  The data and models are
split into four wavelength intervals (defined in the following
section) and, within each interval the spectra are normalized by a
high-order polynomial \citep[with degree $n$ where $n\equiv
(\lambda_{\rm max}-\lambda_{\rm min})/100$\AA; see][for
details]{Conroy12b}.  We have masked the spectral regions surrounding
the H$\alpha$, [\ionn{N}{ii}], [\ionn{S}{ii}], H$\beta$,
[\ionn{O}{iii}], and [\ionn{N}{i}] emission lines.

The effective resolution of the data varies from $R\approx400-1200$
owing to intrinsic Doppler broadening.  Both because of this low
resolution and the fact that in our model the synthetic spectra are
only used differentially, several uncertainties associated with
abundance analysis at high resolution (including the adopted
microturbulent velocity, treatment of convection, non-LTE effects, and
hyperfine splitting of certain levels) are mitigated to some extent
herein.  Moreover, \citet{Bergemann12} has demonstrated that the
strong \ionn{Sr}{ii} line at 4077\AA\, suffers from negligible non-LTE
corrections at [Fe/H]$\approx0.0$.  Of course when working at low
resolution one is more dependent on the fidelity of the model than the
case where equivalent widths of individual unsaturated lines are
directly measured from the data.

As an example of what can be measured from low resolution spectra, in
Figure \ref{fig:response} we show the change in the model spectrum due
to a 0.3 dex increase in the [Sr/Fe] and [Ba/Fe] abundance ratios
(holding [Fe/H] fixed to the solar value).  The reference model is for
an age of 13 Gyr, with a Galactic \citep{Kroupa01} IMF, and for solar
metallicity.  The model was convolved with a velocity dispersion of
$350\kms$.  Notice that even at this low resolution, at least two
\ionn{Sr}{ii} lines and two \ionn{Ba}{ii} lines produce a change in
the spectrum of $>0.3$\% per 0.3 dex change in abundance.  As we will
demonstrate in Section \ref{s:res}, the existence of multiple,
well-separated Sr and Ba lines allows us to test the robustness of our
results in a relatively model-independent fashion.

%-----------------------------------------------------------------%

\section{Data}
\label{s:data}

\begin{figure}[!t]
\center
\resizebox{3.7in}{!}{\includegraphics{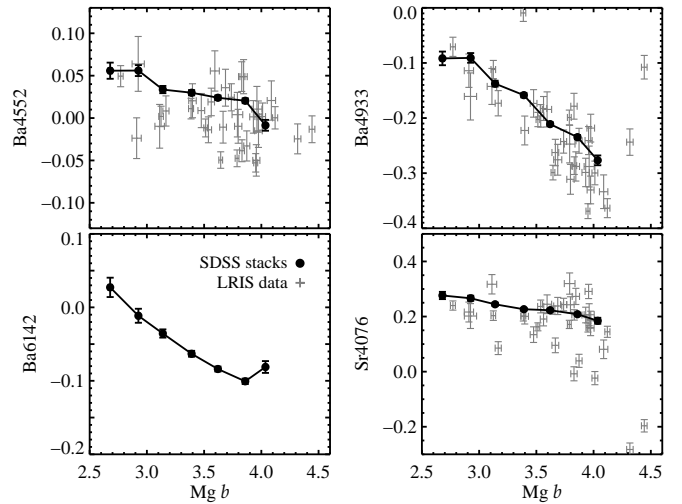}}
\caption{Spectral indices for the stacked SDSS early-type galaxy
  spectra and the LRIS spectra of 35 individual galaxies.  Index units
  are equivalent widths in \AA.  The LRIS data do not cover the Ba6142
  spectral feature.  The Mg\,{\it b} Lick index is sensitive to Mg,
  while the other features are sensitive to Ba and Sr and are defined
  in \citet{Serven05}.}
\label{fig:indx}
\end{figure}

In this Letter we consider two samples of early-type galaxies.  The
first sample of 34 early-type galaxies (and the nuclear bulge of M31)
was presented in \citet{vanDokkum12}.  The sample was selected from
the SAURON survey of nearby galaxies \citep{Bacon01}.  High S/N
spectra of these objects were obtained with the Low Resolution Imaging
Spectrometer (LRIS) on the Keck I telescope.  The wavelength coverage
spans the range $3500$\AA$-10,400$\AA\, with a gap at
$5600$\AA$-7100$\AA.  Four wavelength intervals were defined, $0.4\mu
m-0.46\mu m$, $0.46\mu m-0.55\mu m$, $0.80\mu m-0.89\mu m$, and
$0.96\mu m-1.02\mu m$ for fitting models to data.  The wavelength
choices reflect the limitations of the model and data wavelength
coverage and regions of severe telluric corrections.  The S/N at
5000\AA\, varies from $\approx150$ \AA$^{-1}$ to $\approx450$
\AA$^{-1}$.  Spectra were extracted within $R<R_e/8$, with a weighting
intended to mimic a circular aperture, where $R_e$ is the effective
radius.  These data have already been compared to the CvD12 models in
\citet[][see the Appendix of that paper for a comparison of the
best-fit models and data for each of the 35 LRIS galaxies]{Conroy12b}.
In that earlier work, we did not consider variation in Sr and Ba, and
only briefly commented on the abundances of the other elements
considered in the fit.  We found that the ages, [Fe/H] and [Mg/Fe]
values derived with our full spectrum fitting technique agreed with
conventional Lick index-based techniques.

The second source of data is provided by the Sloan Digital Sky Survey
\citep[SDSS;][]{York00}.  Following the methodology outlined in
\citet{Graves09a}, we have selected a sample of galaxies within a
narrow redshift interval of $0.02-0.06$, with no detected emission in
H$\alpha$ nor in [\ionn{O}{ii}], and with concentrated, de
Vaucouleurs-like light profiles. The SDSS obtains spectra with fibers
that have 3'' diameters.  The typical S/N of SDSS spectra in our
sample is modest, of order 20 \AA$^{-1}$, and so we have chosen to
stack the spectra in seven bins of velocity dispersion with mean
dispersions in each bin of 88, 112, 138, 167, 203, 246, and 300$\kms$.
For the smallest sigma bin the SDSS fiber samples the inner
$0.8\,R_e$, while for the most massive bin the fiber samples the inner
$0.4\,R_e$.  We have verified that our results do not change if we
select galaxies within each bin such that the fiber samples the same
fraction of $R_e$.  Each spectrum was continuum-normalized and
convolved to an effective dispersion of $350\kms$ before stacking, and
each spectrum contributed equally to the stack.  The resulting S/N of
the stacked spectra at 5000\AA\, ranges from $\approx500$ \AA$^{-1}$
to $\approx1800$ \AA$^{-1}$. As with the LRIS data, four wavelength
intervals were defined for the SDSS stacks, $0.4\mu m-0.48\mu m$,
$0.48\mu m-0.58\mu m$, $0.58\mu m-0.64\mu m$, and $0.80\mu m-0.88\mu
m$ for fitting models to data.  These stacked data will be the focus
of future work aimed at measuring the detailed abundance patterns as a
function of galaxy properties.  In this work we limit our attention to
the neutron capture elements Sr and Ba.

As a first look at the data, in Figure \ref{fig:indx} we show spectral
indices for both the SDSS stacked spectra and the individual LRIS
galaxies.  The Mg\,{\it b} index is sensitive to Mg, while the other
indices are sensitive to Ba and Sr lines (with central wavelengths
indicated by the index name).  The Br and Sr indices are defined in
\citet{Serven05}, while the Mg\,{\it b} index is defined in
\citet[][no effort was made to place the latter on the Lick index
scale]{Worthey94b}.  For this figure, the LRIS data were convolved to
a common dispersion of $350\kms$ in order to afford a direct
comparison with the SDSS stacked spectra.  The LRIS data do not cover
the 6142\AA\, spectral region.  Overall the agreement between the LRIS
and SDSS samples is encouraging, given the very different
observational setups and reduction techniques.  It is however
difficult to interpret spectral indices because an index is in general
sensitive not only to the feature of interest but also to features
contributing to the sidebands (i.e., the index measures a feature
strength relative to a pseudocontinuum).  This is especially true when
the feature of interest is weak compared to other features.  Full
spectrum fitting such as the kind performed in the present work does
not suffer from this drawback.

\begin{figure}[!t]
\center
\resizebox{3.5in}{!}{\includegraphics{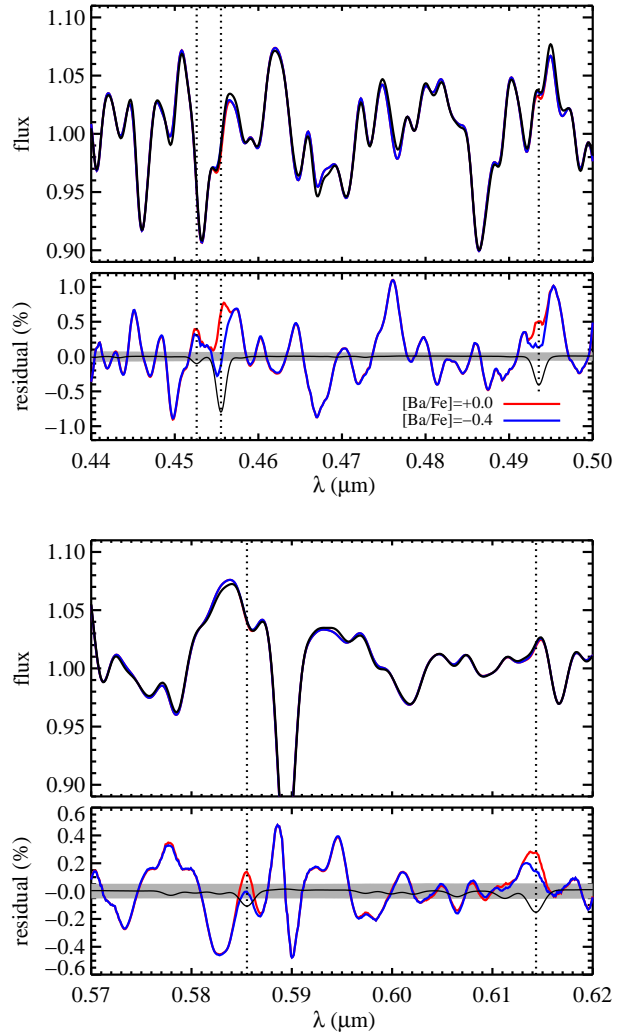}}
\caption{Comparison between models and data in two wavelength regions.
  Upper panels show the continuum normalized spectra and lower panels
  show residuals in percent.  The data (black lines) are for the
  stacked SDSS early-type galaxy spectrum in the dispersion bin
  centered on $\sigma=250\kms$.  Two models are compared, one in which
  the [Ba/Fe] ratio is fixed to the solar value (red line), and a
  model in which the [Ba/Fe] value is included in the fit (with a
  resulting best-fit value of $-0.4$; blue line).  The grey band
  denotes the S/N, and the dotted lines mark the strongest
  \ionn{Ba}{ii} lines in this wavelength range.  The thin line is the
  [Ba/Fe] response function (cf. Figure \ref{fig:response}).  Any
  single \ionn{Ba}{ii} feature on its own is not particularly
  compelling, given the magnitude of the residuals elsewhere in the
  spectrum, but the coincidence of an improvement in $\chi^2$ around
  {\it all four} \ionn{Ba}{ii} lines is strong evidence in favor of a
  sub-solar [Ba/Fe] abundance ratio in this particular spectrum.
  Moreover, the difference in $\chi^2_{\rm min}$ between the two
  models is 1400, signaling strong preference for the model with
  [Ba/Fe]$=-0.4$.}
\label{fig:spec}
\end{figure}

%-----------------------------------------------------------------%

\section{Results}
\label{s:res}

Before showing the derived abundance trends, in Figure \ref{fig:spec}
we explore the quality of the fit around the \ionn{Ba}{ii} lines for
the stacked SDSS spectrum in the second highest $\sigma$ bin. In the
figure we compare two model fits, one in which [Ba/Fe]$=+0.0$ and
another in which the Ba abundance is allowed to vary in the fit (with
a resulting best-fit value of [Ba/Fe]$=-0.4$).  We also mark the
locations of the strongest \ionn{Ba}{ii} lines in this spectral
region.  It is noteworthy that the residuals decrease in {\it all}
spectral regions sensitive to Ba.  This is an important test of the
results to follow because it is also evident from this figure that
there are numerous places where the residuals exceed the formal S/N
limits of the data (demarcated by the grey band).  If there were only
one strong Ba line then it would have been difficult to argue that a
region of high residual required non-solar [Ba/Fe] values.  We return
to this point later in this section.

\begin{figure}[!t]
\center
\resizebox{3.5in}{!}{\includegraphics{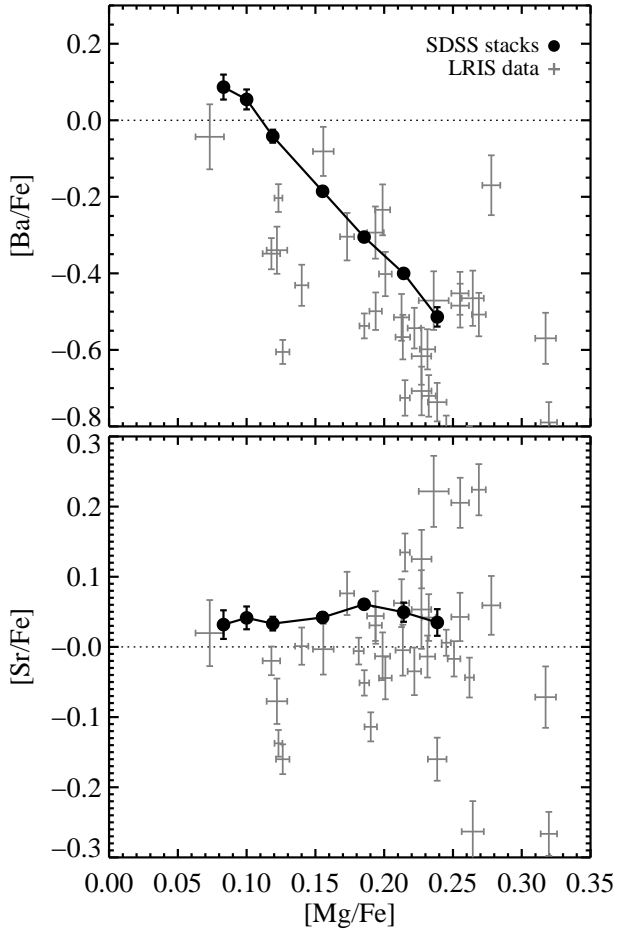}}
\caption{Derived [Ba/Fe] and [Sr/Fe] abundance ratios as a function of
  the derived [Mg/Fe] abundance ratio.  Results are shown both for
  stacked SDSS early-type galaxy spectra and for the LRIS spectra of
  35 individual galaxies.  For the SDSS spectra, the mean velocity
  dispersion increases monotonically with increasing [Mg/Fe].}
\label{fig:basr}
\end{figure}

Our main result is shown in Figure \ref{fig:basr}.  In this figure we
show the best-fit [Ba/Fe] and [Sr/Fe] abundance ratios as a function
of [Mg/Fe].  We include both the stacked SDSS spectra and the LRIS
spectra of individual galaxies.  The formal statistical errors from
the SDSS stacks are very small owing to the very high S/N of the
spectra.  It is encouraging that the LRIS data and SDSS stacks follow
the same general trends of decreasing [Ba/Fe] and constant [Sr/Fe]
with increasing [Mg/Fe].  Such low values of [Ba/Fe] are not uncommon
amongst metal-poor halo stars in the Galaxy \citep{Sneden08}.  Indeed,
below [Fe/H]$\approx-3$ essentially all halo stars have [Ba/Fe]$<0.0$,
with many stars having Ba abundances as low as [Ba/Fe]$\approx-2.0$
\citep{Francois07}.  Such stars are also $\alpha$ enhanced, and have
[Sr/Ba]$>0.0$, again broadly consistent with the mean abundance ratios
of the massive early-type galaxies (although with very different
overall metallicities).

Taken at face value, the [Ba/Fe] versus [Mg/Fe] trends are consistent
with the standard interpretation of [Mg/Fe] probing the SF timescale
in these galaxies.  A shorter timescale should lead to higher [Mg/Fe]
and lower [Ba/Fe] values, as observed.  Although not shown, the SDSS
data also display a strong correlation between [Mg/Ba] and $\sigma$.
The derived [Fe/H] values range from $-0.1$ to $0.0$ over the SDSS
early-type sample, and so the trends observed in Figure \ref{fig:basr}
are unlikely to be due to metallicity-dependent yields.  Other
explanations for the [Mg/Fe]$-\sigma$ relation have been proposed,
including a varying IMF, varying SN Ia rate, and selective mass-loss
of SN Ia ejecta \citep{Worthey92, Thomas99, Trager00}.  The [Mg/Ba]
data allow us to rule out the explanations that are particular to the
Ia's, since the Ia's influence neither Mg nor Ba, but it cannot
discriminate between a varying SF timescale and a varying high-mass
IMF for the observed variation in [Mg/Fe] with $\sigma$.

The near constancy of [Sr/Fe] is puzzling.  It has been suggested that
Sr may have multiple nucleosynthetic origins, including the
$r$-process, both the weak and main $s$-processes, and perhaps even
hypernovae \citep{Raiteri93, Travaglio04, Qian08}.  As [Mg/Fe] (and
presumably the SF timescale) varies, then perhaps the main
nucleosynthetic site varies in such a way as to keep a rough constancy
in [Sr/Fe].  With detailed chemical evolution models constrained to
match the observed [Mg/Fe] and [Ba/Fe] trends, it may be possible to
place interesting constraints on the origin(s) of Sr in these massive
galaxies.

\begin{figure}[!t]
\center
\resizebox{3.5in}{!}{\includegraphics{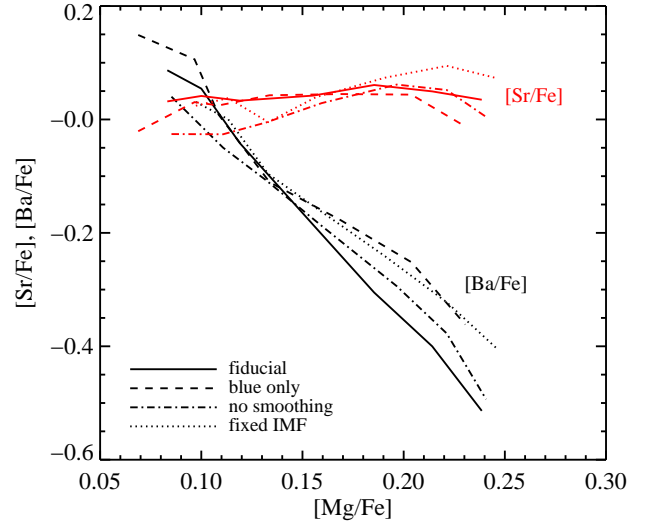}}
\caption{Exploration of systematic effects on the [Sr/Fe] and [Ba/Fe]
  abundance trends.  The fiducial model is compared to models in which
  only the blue ($\lambda<5800$\AA) spectral region is fit, in which
  there is no smoothing of the stacked spectra to $\sigma=350\kms$,
  and in which the IMF is fixed to the Galactic form.  Results are
  shown only for the stacked SDSS early-type galaxy spectra for
  clarity.  While these model variations induce changes of the order
  of 0.1 dex, the trends are clearly robust to these model
  permutations.}
\label{fig:sys}
\end{figure}

We turn now to several tests that have been performed to assess the
robustness of these results.  In Figure \ref{fig:sys} we show the same
abundance trends as before, for only the SDSS stacked spectra, and for
a variety of permutations to the fiducial model.  We consider a model
in which the IMF is held fixed to the Galactic \citep{Kroupa01} form,
a model in which only the blue ($\lambda<5800$\AA) spectral region is
fit, and a model in which the SDSS stacks are not smoothed to a common
resolution of $\sigma=350\kms$.  The latter is a particularly
important test because much more information is available in the low
dispersion bins, which effectively have higher spectral resolution,
compared to the high dispersion bins. Relatedly, when the data are not
broadened to $\sigma=350\kms$, the associated response functions
(Figure \ref{fig:response}) become much stronger.  The derived
abundance ratios vary between these model permutations by $\sim0.1$
dex, but the overall qualitative trends are clearly robust to these
details.  We have also masked each of the two strongest \ionn{Ba}{ii}
lines (one at a time) and refit the models to the data.  The resulting
[Ba/Fe] versus [Mg/Fe] trends are unaltered, although in the case of
masking the \ionn{Ba}{ii} 4555\AA\, line the overall [Ba/Fe]
abundances shift lower by $\approx0.2$ dex.  This highlights that
while trends appear to be robust in our analysis, there are zero-point
uncertainties at the $0.1-0.2$ dex level.

A concern is that the features we are trying to measure are much
weaker than other metal lines. In fact, as shown in Figure
\ref{fig:spec} the expected Ba line strengths are similar to the
typical systematic residual in the spectra {\em after} subtraction of
our best-fit model!  A robust test of our result is to randomly
shuffle the central wavelengths of the strongest \ionn{Ba}{ii} lines
and recompute $\chi^2$ around the shuffled lines.  In practice, this
is achieved by altering the response function for Ba (see Figure 1) by
assigning random central wavelengths in the interval
$4000$\AA$-6000$\AA\, while keeping the relative strengths and widths
of the lines the same.  We have performed this test $10^3$ times for
each of the SDSS stacked spectra.  As an example of the results for
the highest $\sigma$ bin, the residuals decrease around the strongest
shuffled \ionn{Ba}{ii} line in only $0.1$\% of the shuffled models
when using the best-fit [Ba/Fe] value shown in Figure \ref{fig:basr}.
Moreover, in {\it none} of the shuffled models do the residuals
decrease in both of the two strongest shuffled \ionn{Ba}{ii} lines.
In the four highest $\sigma$ bins, where the best-fit [Ba/Fe] ratios
are $<-0.1$, the shuffled models are never able to reduce the
residuals in all three of the strongest \ionn{Ba}{ii} lines.  It is
therefore highly unlikely that the derived [Ba/Fe] abundances are the
result of a chance alignment between the strong \ionn{Ba}{ii} lines
and spectral regions where the residuals happen to be high.

The elemental abundances derived herein should provide novel
constraints on the duration of star formation in early-type galaxies
once chemical evolution models that include yields from the $s$- and
$r$-processes are employed.  The [Mg/Ba] ratio is particularly
promising because the derived timescale will not depend on the
uncertain delay time distribution or overall rate of Type Ia SNe, in
contrast to timescales derived from the [Mg/Fe] ratio.  In future work,
we will present detailed abundance patterns for $\approx17$ elements,
which promise to provide further insight into the early formation
histories of these now dormant galaxies.

%-----------------------------------------------------------------%

\acknowledgments 

Some of the data presented herein were obtained at the W.M. Keck
Observatory, which is operated as a scientific partnership among the
California Institute of Technology, the University of California and
the National Aeronautics and Space Administration. The Observatory was
made possible by the generous financial support of the W.M. Keck
Foundation.  Funding for the SDSS and SDSS-II has been provided by the
Alfred P. Sloan Foundation, the Participating Institutions, the
National Science Foundation, the U.S. Department of Energy, the
National Aeronautics and Space Administration, the Japanese
Monbukagakusho, the Max Planck Society, and the Higher Education
Funding Council for England. The SDSS Web Site is
http://www.sdss.org/.  The SDSS is managed by the Astrophysical
Research Consortium for the Participating Institutions. The
Participating Institutions are the American Museum of Natural History,
Astrophysical Institute Potsdam, University of Basel, University of
Cambridge, Case Western Reserve University, University of Chicago,
Drexel University, Fermilab, the Institute for Advanced Study, the
Japan Participation Group, Johns Hopkins University, the Joint
Institute for Nuclear Astrophysics, the Kavli Institute for Particle
Astrophysics and Cosmology, the Korean Scientist Group, the Chinese
Academy of Sciences (LAMOST), Los Alamos National Laboratory, the
Max-Planck-Institute for Astronomy (MPIA), the Max-Planck-Institute
for Astrophysics (MPA), New Mexico State University, Ohio State
University, University of Pittsburgh, University of Portsmouth,
Princeton University, the United States Naval Observatory, and the
University of Washington.

%\bibliography{../master_refs}

\end{document}